# Theory of activated-rate processes under shear with application to shear-induced aggregation of colloids

Alessio Zaccone, Hua Wu, Daniele Gentili, and Massimo Morbidelli

Department of Chemistry and Applied Biosciences

ETH Zurich, 8093 Zurich, Switzerland

CORRESPONDING AUTHOR

Alessio Zaccone

Email: alessio.zaccone@chem.ethz.ch.

Fax: 0041-44-6321082.

# ABSTRACT

Using a novel approximation scheme within the convective diffusion (two-body Smoluchowski) equation framework, we unveil the shear-driven aggregation mechanism at the origin of structure-formation in sheared colloidal systems. The theory, verified against numerics and experiments, explains the induction time followed by explosive (irreversible) rise of viscosity observed in charge-stabilized colloidal and protein systems under steady shear. The Arrhenius-type equation with shear derived here, extending Kramers' theory in the presence of shear, is the first analytical result clearly showing the important role of shear-drive in activated-rate processes as they are encountered in soft condensed matter.

# I. INTRODUCTION

Complex and biological fluids display intriguing rheological properties, depending on the nature and interactions of their microscopic constituents (colloidal and noncolloidal particles such as cells, biological and synthetic macromolecules, inorganic particulates, nanoparticles etc) [1]. Under steady shear, *thixotropy* (where viscosity decreases with time) and its opposite, *rheopexy* or *anti-thixotropy* (where viscosity increases with time), are ubiquitous and call for a deeper understanding of the interplay between structure-breaking and structure-building processes from which they originate [2,3]. In particular, rheopexy as due to aggregation under shear, is observed in many biological fluids such as protein solutions (e.g. the synovial fluid which lubricates mammalian freely moving joints) and blood plasmas [4,5]. Rheopexy, under certain conditions, may end into a liquid-solid (jamming) transition in time which attracts considerable attention for the fabrication of new materials with unusual properties. It has been shown, for instance, that rheopexy accompanies the formation of spider silk within the spider's spinneret, where high shear rates induce the formation of large aggregates that jam into a light material with formidable mechanical response [5,6]. Nevertheless, the current understanding is qualitative and largely built upon empirical evidence. It is not clear why, under most conditions, viscosity increases sharply after an induction period (during which it is equal to the zero-time viscosity), reminiscent of an explosive behaviour [2,5,7]. The induction period, typically observed in systems with a repulsive barrier (e.g. charge-stabilization), is highly suggestive of an activation mechanism for irreversible aggregation driven by shear, as was recently speculated in [7]. However, due to the nontrivial interplay between shear, direct (interparticle) interactions, and Brownian motion, no clear understanding of such phenomena is currently at hand. We thus propose a novel theory based upon the convective diffusion equation or two-body Smoluschowski equation with shear. The results unveil the nature of shear-driven activation which triggers self-accelerating (irreversible) aggregation kinetics once that an activated (cluster) size is reached. Our theory, whose qualitative and quantitative predictions are verified against both experiments and numerics, explains the main features observed in the steady-shear rheology of interacting colloids, such as protein solutions, by means of an Arrhenius-type rate-equation accounting for shear-driven activation.

## II. DERIVATION

### A. Two-body Smoluchowski equation with shear

Let us consider a dispersion of diffusing particles interacting with a certain interaction potential. The (number-)concentration field will be called $c(\mathbf{r})$. The associated normalized probability density function $g(\mathbf{r}) \equiv \tilde{c}(\mathbf{r})$ for finding a second particle at distance $\mathbf{r}$ from a reference one is then normalized such that $c(\mathbf{r}) = c_0 \tilde{c}(\mathbf{r})$, where $c_0$ is the bulk concentration. The evolution equation thus reads

$$\partial c / \partial t = \text{div}(D\nabla c - \beta D \mathbf{K} c) \qquad (1)$$

where $D$ is the mutual diffusion coefficient of the particles ($D = 2D_0 \mathcal{G}(r)$, $D_0$ being the diffusion coefficient of an isolated particle and $\mathcal{G}(r)$ the hydrodynamic function for viscous retardation), $\beta = 1/k_B T$ is the Boltzmann factor, and $\mathbf{K}$ is the arbitrary external force. According to this equation, the associated stationary current is given by

$$\mathbf{J} = (\beta D \mathbf{K} - D \nabla) c \qquad (2)$$

Hence, when steady-state is reached, $\partial c(\mathbf{r})/\partial t = 0$ and

$$\text{div}(\beta D\mathbf{K} - D\nabla)c = 0 \tag{3}$$

If the dispersing medium is subjected to an externally applied flow, the arbitrary external force field for our problem may be decomposed into two terms, a non-conservative one accounting for the drift caused by the flow velocity $\mathbf{v}(\mathbf{r}) = [v_r, v_\theta, v_\varphi]$, and the term accounting for the conservative force field due to the two-body direct interaction between particles $-\nabla U(r)$. Introducing the hydrodynamic drag $b = 3\pi\mu a$, one obtains

$$\mathbf{K}(\mathbf{r}) = -\nabla U(r) + b\mathbf{v}(\mathbf{r}). \tag{4}$$

where $\mu$ is the viscosity of the solvent and $a$ is the particle radius. Hence, the following two-particle Smoluchowski equation with shear can be written as

$$\text{div}\left[\beta D(-\nabla U + b\mathbf{v}) - D\nabla\right]c = 0 \tag{5}$$

with the associated current given by

$$\mathbf{J} = \left[\beta D(-\nabla U + b\mathbf{v}) - D\nabla\right]c \tag{6}$$

We can now define the collision frequency or collision rate across a spherical surface of radius $r$, concentric with the stationary particle, as

$$\begin{aligned} G = \oint \mathbf{J} \cdot \mathbf{n}\, dS &= \oint \left[D\nabla c + \beta D(\nabla U - b\mathbf{v})c\right] \cdot \hat{\mathbf{n}}\, dS \\ &= 4\pi r^2 D\left[\left(\frac{\partial \langle c\rangle}{\partial r} + \beta \frac{\partial U}{\partial r}\langle c\rangle\right)\right] + 4\pi r^2 b\langle v_r^+ c\rangle \end{aligned} \tag{7}$$

where $dS = r^2 \sin\theta\, d\theta\, d\phi$ denotes the element of (spherical) collision surface, while $\langle...\rangle$ denotes the isotropic average, $(4\pi)^{-1}\int_0^{2\pi} d\varphi \int_0^\pi \sin\theta\, d\theta$. Recall that $G$, by definition, is the *inward* flux of particles through the spherical surface [8,9]. Therefore, integration runs exclusively over those orientations (or, equivalently, those pairs of angles $\theta$, $\varphi$) such that $\mathbf{v} \cdot \hat{\mathbf{n}} = -v_r$, which define the upstream region. Hence we will use $v_r^+(\mathbf{r})$ to denote the positive part of the radial component of the fluid velocity, $v_r(\mathbf{r})$. Thus

$$v_r^+(\mathbf{r}) = \max(v_r(\mathbf{r}), 0) = \begin{cases} v_r(\mathbf{r}) & \text{if } v_r(\mathbf{r}) > 0 \\ 0 & \text{else} \end{cases} \quad (8)$$

We use the approximation that the flow velocity and the concentration profile around the stationary particle are weakly correlated in space, which is expressed through

$$\langle (v_r^+(\mathbf{r}) - \langle v_r^+(\mathbf{r}) \rangle)(c(\mathbf{r}) - \langle c(\mathbf{r}) \rangle) \rangle \approx 0, \quad (9)$$

thus leaving $\langle v_r^+(\mathbf{r}) c(\mathbf{r}) \rangle \approx \langle v_r^+(\mathbf{r}) \rangle \langle c(\mathbf{r}) \rangle$. This approximation is expected to do a reasonably good job as long as the Brownian motion is effective in randomizing the particle concentration, i.e. when Brownian motion is not overwhelmed by convection. Eq. (7), under this approximation, simplifies to

$$G = 4\pi r^2 D \left( \frac{\partial}{\partial r} + \beta \frac{\partial U}{\partial r} + b \langle v_r^+ \rangle \right) \langle c \rangle \quad (10)$$

Clearly, the collision rate given by Eq. (10) corresponds to the collision rate that we would have if the actual flow field $\mathbf{v}(\mathbf{r})$ were replaced by an *effective* flow field for aggregation where only the radial component, given by $\langle v_r^+(\mathbf{r}) \rangle$, is non-zero. The effective flow field will be denoted as $\mathbf{v}_{\text{eff}}(r) = [v_{r,\text{eff}}(r), 0, 0]$ with $v_{r,\text{eff}}(r) \equiv \langle v_r^+(\mathbf{r}) \rangle$. Therefore, a system where the collision rate (hence the colloidal aggregative stability) is identical to the one of the real system may be described, under the approximation Eq. (9), by the following *effective* two-particle Smoluchowski equation for the orientation-averaged concentration field $\langle c \rangle$

$$\text{div}\left[\beta D\left(\nabla U + b\mathbf{v}_{r,\text{eff}}\right) + D\nabla\right]\langle c\rangle = 0 \tag{11}$$

which, for isotropic potential, can be written as an ordinary differential equation

$$\frac{1}{r^2}\frac{d}{dr}\left[r^2 D\left(\beta\frac{dU}{dr} + b\mathbf{v}_{r,\text{eff}}\right)\langle c\rangle + Dr^2\frac{d\langle c\rangle}{dr}\right] = 0 \tag{12}$$

**B. The far-field boundary condition with linear flow fields: boundary-layer analysis**

The boundary conditions for the irreversible aggregation problem are as follows. First, the reaction kinetics by which the particle irreversibly sticks to the reference one is taken to be infinitely fast at $r = 2a$, which corresponds to the familiar absorbing boundary condition $\langle c\rangle = 0$ at $r = 2a$ [8]. Second, the bulk concentration ($c/c_0 = 1$) must be recovered at a certain distance from the reference particle. This condition is often implemented at large distance, namely at $r \to \infty$, which is always possible for velocity fields which decay to zero at infinity. Classic examples are the problems of convective diffusion of a solute to a free-falling particle or convective diffusion to a rotating disk (cfr. Levich [8], Ch. 2). However, it is well known that in the case of *linear* velocity fields, application of the second (far field) boundary condition, $c/c_0 = 1$ at $r \to \infty$, is more complicated, due to a singularity at the domain boundary induced by $\mathbf{v} = \mathbf{\Gamma}\cdot\mathbf{r}$, where $\mathbf{\Gamma}$ is the velocity gradient tensor. In this case, the Smoluchowski equation becomes

$$\text{div}\left[\left(-\beta D\frac{\partial U}{\partial r} + \mathbf{\Gamma}\cdot\mathbf{r}\right) - D\nabla\right]c = 0 \tag{13}$$

As first diagnosed by Dhont [10] who used Eq. (13) to study the structural distortion of sheared non-aggregating suspensions [11], the $\mathbf{\Gamma}\cdot\mathbf{r}$ term, being linear in $\mathbf{r}$, overwhelms the other terms at sufficiently large separations, even for very small shear rates. It was shown that in the case of hard spheres the extent of separation $\delta$ where this occurs decreases with the Peclet number as $\delta/a \sim Pe^{-1/2}$ [10]. In other terms, $\delta$ defines a boundary-layer width beyond which convection represents by far the controlling phenomenon [8,10]. To overcome the difficulty of having a singular term at the domain boundary which makes it impossible to enforce the far-field boundary-condition, specific considerations are required. For example, when the final goal is to determine the structure factor of a suspension, it is convenient to Fourier-transform the radial domain or equivalently to move to a reciprocal domain $q = 2/r$ (see e.g. [12]). Alternatively, within numerical studies in real space, the far-field boundary condition is usually applied at finite separations, after self-consistently identifying the location beyond which the concentration profile flattens out as a consequence of convection becoming predominant [13,14].

At this point, it is useful to consider the boundary-layer structure of the Smoluchowski equation and of the formally identical convective diffusion equation. In fact, as is well known within the theory of convective diffusion [8], due to the boundary-layer behaviour of Eq. (13), the convective flux dominates at sufficiently large interparticle separations (which has been verified numerically in [13]) since the other terms in the bracket in Eq. (13) become negligible in comparison with $\mathbf{\Gamma} \cdot \mathbf{r}$. That region of space is thus described by

$$\text{div}(\mathbf{v}c) = 0 \tag{14}$$

which admits the solution $c = const$ [8]. For colloidal particles in a linear flow field, Eq. (14) has been solved under account of hydrodynamic interactions by Batchelor and Green [15] who derived the following solution in terms of the pair-correlation function

$$g(r) = \frac{1}{1-A} \exp \int_r^\infty dr \frac{3(B-A)}{r(1-A)} \tag{15}$$

where the hydrodynamic functions $A(r)$ and $B(r)$ describe the nonlinear disturbance to the relative velocity due to the particle motion. Beyond a few particle radii distance both $A(r)$ and $B(r)$ decay to zero, and we have $g(r) = 1$ thus implying $c = const = c_0$. Hence, the mere effect of *convection* in a linear flow-field, in the absence of any perturbation (as could arise from diffusive motion or hydrodynamic interactions) is to *flatten out* the concentration profile. Therefore, if it is true that *convection* prevails at separations larger than the boundary-layer width, it must follow that assuming homogeneity, $c = const = c_0$, at separations larger than the boundary-layer is justified. In this work we thus propose using $r = \delta + 2a$ (instead of $r \to \infty$) in the far-field boundary condition. In the language of matched asymptotic expansions, this corresponds to exactly determining the solution within the boundary-layer and matching it to the leading order in $Pe^{-1}$ expansion in the outer layer. (Note however that the problem of Eq. (13) in real space is substantially more complicated than standard singularly perturbed equations due to the aforementioned singular behaviour of $\mathbf{\Gamma} \cdot \mathbf{r}$ at the domain boundary, in addition to the standard singular behaviour for large values of the Peclet number, as first noted by Dhont [10].) Hence, the collision kinetics being uniquely determined by the inner solution, the only approximation involved is on the location of the far-field boundary-condition, which we estimate in the following as a function of Peclet number and interaction parameters. The good accuracy of the estimate is subsequently verified by quantitative comparison with numerics.

The width of the boundary-layer $\delta$ where the major change in the concentration profile occurs, and within which diffusion, convection and direct interactions are all important, can be uniquely determined from dimensional considerations. The governing parameters upon which $\delta$ depends are: $\lambda$, $a$, $\dot{\gamma}$ and $D_0$, where $\lambda$ denotes the range of interaction. Thus the dimensionless boundary-layer width $\delta/a$ is expressible as a dimensionless combination of the governing parameters (Rayleigh)

$$\delta/a \sim \lambda^l a^m \dot{\gamma}^n D_0^p \tag{16}$$

From the boundary-layer behaviour of the Smoluchowski equation for hard spheres, it is known that $\delta/a \sim Pe^{-1/2}$ [10], which identifies $n = -1/2$ and $p = 1/2$. Hence considering that $[\lambda] = [a] = L$ and $[\dot{\gamma}]^{-1/2}[D_0]^{1/2} = L$, it obviously follows

$$l + m + 1 = 0 \tag{17}$$

Application of the $\Pi$-theorem of dimensional analysis prescribes that $\delta/a$ has to be expressed in terms of two independent dimensionless groups, $\lambda/a$ and $\dot{\gamma}a^2/D_0$, as $\delta = a\Phi(\lambda/a, \dot{\gamma}a^2/D_0)$, because $a$ and $\dot{\gamma}$ are parameters with independent dimension [16]. This fixes the $m$ value:

$$m = -3/2 \tag{18}$$

It is therefore concluded that the dimensionless boundary-layer width $\delta/a$ is approximately given by

$$\delta/a \sim \sqrt{(\lambda/a)/Pe}. \tag{19}$$

When Eq. (19) is compared to the case of hard spheres, $\delta/a \sim Pe^{-1/2}$, the effect of the colloidal interactions on the boundary-layer width enters through the interaction range $\lambda$, which for the screened-Coulomb repulsion is given by $\lambda = \kappa^{-1}$, where $\kappa^{-1}$ is the Debye length.

### C. Approximate solution for the aggregation rate

Based on the above boundary value problem analysis, let us rewrite the effective two-body Smoluchowki equation in shear, Eq. (12), in dimensionless form [17]

$$0 = \frac{1}{Pe}\frac{1}{(x+2)^2}\frac{d}{dx}(x+2)^2\frac{dC}{dx} + \frac{1}{Pe}\frac{1}{(x+2)^2}\frac{d}{dx}\left((x+2)^2\frac{d\widetilde{U}(x)}{dx}C\right) + \frac{1}{(x+2)^2}\frac{d}{dx}\left[(x+2)^2\tilde{v}_{r,\text{eff}}C\right] \quad (20)$$

with the boundary conditions

$$\begin{aligned} C &= 0 \quad \text{at} \quad x = 0 \\ C &= 1 \quad \text{at} \quad x = \delta/a \end{aligned} \quad (21)$$

where, for simplicity of notation, we have set $\langle \tilde{c}(\mathbf{r}) \rangle \equiv C(x)$, since the latter is a function only of the dimensionless separation $x = (r/a) - 2$. The tilde indicates non-dimensionalized quantities. The Peclet number is given by $Pe = \dot{\gamma}a^2/D = 3\pi\mu\dot{\gamma}a^3/k_BT$. Eq. (20) is formally identical to a stationary one-dimensional Fokker-Planck equation in spherical geometry with time independent drift and diffusion coefficients [18]. The concentration profile in dimensional form after application of the second boundary condition to Eq. (20) reads

$$\begin{aligned} \langle c \rangle &= \left\{ \exp \int_{\delta/a}^{x} dx(-\beta dU/dx - Pe\tilde{v}_{r,\text{eff}}) \right\} \\ &\quad \times \left\{ c_0 + \frac{G}{8\pi a D_0} \int_{\delta/a}^{x} \frac{dx}{\mathcal{G}(x)(x+2)^2} \exp \int_{\delta/a}^{x} dx(\beta dU/dx + Pe\tilde{v}_{r,\text{eff}}) \right\} \end{aligned} \quad (22)$$

The flux $G$ is determined from the absorbing boundary condition at contact as

$$G = \frac{8\pi D_0 a c_0}{2\int_0^{\delta/a} \frac{dx}{\mathcal{G}(x)(x+2)^2} \exp\int_{\delta/a}^{x} dx(\beta dU/dx + Pe\tilde{v}_{r,\text{eff}})} \tag{23}$$

Using $\delta/a = \sqrt{(\lambda/a)/Pe}$, Eq. (23) can be integrated with standard numericals and can find direct application to systems under shear and straining flows, provided that an appropriate form of the velocity field is given. Note that the effect of hydrodynamic viscous dissipation is accounted for through the function $\mathcal{G}(x)$. For an axisymmetric extensional flow, the radial component of the velocity field is given by [13,15]

$$v_r = \frac{1}{2}\dot{\gamma}a(x+2)[1-A_E(x)](3\cos^2\theta - 1) \tag{24}$$

where $A_E(x)$ is the corresponding function accounting for the effect of hydrodynamic retardation. Based on how the rescaled effective velocity has been defined, we thus obtain

$$\tilde{v}_{r,\text{eff}}(x) \equiv \langle \tilde{v}_r^+(\mathbf{r})\rangle = -(1/3\sqrt{3})(x+2)[1-A_E(x)] \tag{25}$$

Similarly, in the case of pure laminar shear we have

$$\tilde{v}_{r,\text{eff}}(x) \equiv \langle \tilde{v}_r^+(\mathbf{r})\rangle = -(1/3\pi)(x+2)[1-A_S(x)] \tag{26}$$

where $A_S(x)$ is the hydrodynamic retardation function for simple shear. Hence, within this formulation it has been possible to account for hydrodynamic interactions (in the two-body limit) resulting from viscous dissipation, through $\mathcal{G}(x)$, and from hydrodynamic disturbance (retardation) induced by the second particle, through $A(x)$.

### D. Irreversible aggregation kinetics and colloidal stability with shear

Since we have retained all terms in the governing equation (and built the solution exactly in the inner layer), Eq. (23) is valid for arbitrary thickness of the boundary layer $\delta$. In particular we observe that in the limit $Pe \to 0$, Eq. (23) reduces to the well-known Fuchs' formula for the aggregation rate constant (collision rate) in the presence of direct interaction forces but in the absence of flow, which reads [9]

$$G = \frac{8\pi D_0 a c_0}{2\int_0^\infty dx \frac{e^{\beta U(x)}}{\mathcal{G}(x)(x+2)^2}} \tag{27}$$

Comparing Eqs. (23) and (27) leads to the definition of a generalized stability coefficient which is valid for arbitrary $Pe$ numbers and interaction potentials

$$W_G = 2 \int_0^{\sqrt{(\lambda/a)/Pe}} \frac{dx}{\mathcal{G}(x)(x+2)^2} \exp \int_{\sqrt{(\lambda/a)/Pe}}^x dx(\beta dU/dx + Pe\tilde{v}_{r,\text{eff}}) \tag{28}$$

This, at $Pe=0$, reduces to the stagnant colloidal stability coefficient first derived by N. A. Fuchs in 1934 (cfr. [9]):

$$W = 2\int_0^\infty dx \frac{e^{\beta U(x)}}{\mathcal{G}(x)(x+2)^2} \tag{29}$$

Thus, the combined effect of fluid motion (convection) and direct interactions can either diminish or augment the coagulation rate with respect to the case of non-interacting Brownian particles in a stagnant fluid by a factor equal to $W_G$.

### E. Initial kinetics of irreversible aggregation

In deriving Eq. (23) for the collision rate against a stationary particle we did not account for the diffusive motion of the latter. Accounting for that leads to a factor two times the macroscopic number concentration of particles in the system, $c$. Hence, the kinetic equation for the rate of change of the concentration of particles reads

$$\frac{dc}{dt} = -\frac{16\pi D_0 a}{W_G} c^2 \qquad (30)$$

where $W_G$ is the generalized stability coefficient given by Eq. (28). Integrating Eq. (30) yields the law of variation with time of the particle concentration

$$c(t) = \frac{c_0}{1 + t/t_c} \qquad (31)$$

where

$$t_c = (16\pi D_0 a c_0 / W_G)^{-1} \qquad (32)$$

is the characteristic time of aggregation. Its reciprocal value defines the kinetic constant for the aggregation of primary particles

$$k_{1,1} = 16\pi D_0 a c_0 / W_G \qquad (33)$$

Since $W_G$, which is given by Eq. (28), brings about a complex dependence upon the particle size when $Pe > 0$, it can be anticipated that the evolution of the aggregation process will substantially differ from the purely Brownian case in a stagnant medium, with important consequences for the dynamics of new-phase formation in sheared colloids.

### F. Comparison with numerical results from the literature

Let us compare the theoretical predictions from Eq. (23) with numerical results where the full convective diffusion equation, Eq. (3), was solved numerically, by means of a finite difference method. The colloidal system is composed of colloidal particles with $a = 100$ nm, with interactions via standard DLVO potential, and convection is induced by laminar axysimmetric extensional flow [13]. The numerically obtained values of

$c(\mathbf{r})$ were then used to determine the aggregation rate constant from numerical evaluation of

$$G = \oint \mathbf{J} \cdot \hat{\mathbf{n}} \, dS = \oint \left[ D\nabla c + \beta D(\nabla U - b\mathbf{v})c \right] \cdot \hat{\mathbf{n}} \, dS \tag{34}$$

where the collision surface is taken as the spherical surface of radius $2a$.

The comparison is shown in Fig. (2) for three different values of ionic strength and a fixed surface potential equal to -14.7 mV. The direct potential $U$ for the same conditions of the numerical simulations, as well as the hydrodynamic functions $A(x)$ and $\mathcal{G}(x)$, have been consistently calculated according to [13]. As shown in the figure, the theory is able to quantitatively reproduce the numerical data for all conditions. In particular, it is seen that the inflection point which marks the transition from a purely-Brownian like regime at $Pe < 1$ to a shear-dominated regime at $Pe > 10$ is very well captured by the theory. Some underestimation arises in the regime of high Peclet numbers $Pe > 50$, which tends to become more important upon further increasing $Pe$. Such underestimation is related to the approximation $\langle (v_r^+(\mathbf{r}) - \langle v_r^+(\mathbf{r}) \rangle)(c(\mathbf{r}) - \langle c(\mathbf{r}) \rangle) \rangle \approx 0$ made in the derivation of Eq. (23). Clearly, the spatial correlation between the flow velocity and the concentration field around the stationary particle would become non-negligible at high Peclet numbers. In this regime, in fact, the randomizing effect of Brownian motion is progressively lost, whereas the angular regions (relative orientations between particles) where the flow velocity is higherand inwardly-directed tend to coincide with the regions where the probability of finding incoming particles is higher.

## G. Potential barrier crossing as an activated-rate process enhanced by shear

Let us consider the case of a high potential barrier in the interaction between particles (as in the case of charge-stabilized colloids at low ionic strength, as predicted by DLVO theory). Further, we will neglect the effect of the hydrodynamic retardation on the velocity field., $A(x) = 0$ in Eqs. (25)-(26), so that the effective velocity reads $\tilde{v}_{r,\text{eff}} = -\alpha(x+2)$, where $\alpha$ is a numerical coefficient which depends upon the type of flow (e.g. $\alpha = 1/3\pi$ for simple shear). Then, the integrand in the second integral in the denominator of Eq. (23) reduces to

$$\int_{\sqrt{(\lambda/a)/Pe}}^{x} dx(\beta dU/dx + Pe\tilde{v}_{r,\text{eff}}) \approx \beta U - \beta U\big|_{x=\sqrt{(\lambda/a)/Pe}} - \frac{\alpha}{2}Pe(x+2)^2 + \frac{\alpha}{2}\lambda \tag{35}$$

When $Pe$ is not too high, $U\big|_{x=\sqrt{(\lambda/a)/Pe}} \approx 0$. Thus the denominator on the r.h.s. of Eq. (23) becomes

$$2\int_{0}^{\sqrt{(\lambda/a)/Pe}} \frac{dx}{\mathcal{G}(x+2)^2} \exp \int_{\sqrt{(\lambda/a)/Pe}}^{x} dx(\beta dU/dx + Pe\tilde{v}_{r,\text{eff}})$$

$$\approx 2e^{\alpha\lambda/2 - \beta U|_{x=\sqrt{\lambda/Pe}}} \int_{0}^{\sqrt{(\lambda/a)/Pe}} \frac{dx}{\mathcal{G}(x+2)^2} \exp\left[\beta U - \alpha Pe(x+2)^2/2\right] \tag{36}$$

Since $U$ goes through an interaction-potential maximum (barrier) in $x \in [0, \sqrt{(\lambda/a)/Pe}]$, so does the function $\beta U - \alpha Pe(x+2)^2/2$. The argument of the exponential can thus be expanded near the maximum up to second order

$$\beta U - \alpha Pe(x+2)^2/2 \approx \beta U_m - \alpha Pe(x_m+2)^2/2 + (U_m'' - \alpha Pe)(x-x_m)^2 \tag{37}$$

where the subscript $m$ indicates quantities evaluated at the potential maximum. We can thus evaluate the remaining integral

$$\int_0^{\sqrt{(\lambda/a)/Pe}} \frac{dx}{\mathcal{G}(x)(x+2)^2} \exp\left[(\beta U_m'' - \alpha Pe)(x-x_m)^2\right] \tag{38}$$

by the steepest descent (Laplace) method to finally obtain

$$W_G \approx \sqrt{\frac{2\pi}{\alpha Pe - \beta U_m''}} \frac{2e^{\alpha\lambda/2 - \beta U|_{x=\sqrt{(\lambda/a)/Pe}}}}{(x_m+2)^2 \mathcal{G}(x_m)} e^{\beta U_m - \alpha Pe(x_m+2)^2/2} \tag{39}$$

More precise approximations can be obtained by considering higher order terms in the expansion Eq. (37) [18]. From Eq. (39), in view of being $x_m + 2 \approx 2$, the effect of the Peclet number and potential barrier on the two-body aggregation rate constant, $k_{1,1}$, is given by

$$k_{1,1} \sim \sqrt{\alpha Pe - \beta U_m''} e^{-\beta U_m + 2\alpha Pe} = \sqrt{\frac{3\pi\alpha\mu\dot{\gamma}a^3 - U_m''}{k_B T}} e^{-(U_m - 6\pi\alpha\mu\dot{\gamma}a^3)/k_B T} \tag{40}$$

It is interesting to observe that Eq. (40) appears to be in Arrhenius form, with the pre-exponential or frequency factor $\sqrt{(3\pi\alpha\mu\dot{\gamma}a^3 - U_m'')/k_B T}$, and the activation energy $U_m - 6\pi\alpha\mu\dot{\gamma}a^3$. Note that due to $U_m$ being the potential maximum, $U_m'' < 0$. In both parameters the shear rate $\dot{\gamma}$ plays a prominent role. Increasing $\dot{\gamma}$ leads to increasing the collision rate, through the prefactor, and at the same time to decreasing the activation energy (thus increasing the fraction of successful collisions). Of course, for substantially large $\dot{\gamma}$ values, $\dot{\gamma}$ has a dominant effect on the aggregation rate due to the exponential form. Further, a critical value of the shear rate can be defined, which corresponds to vanishing activation energy,

$$\dot{\gamma}^* = U_m / 6\pi\alpha\mu a^3 \tag{41}$$

When $\dot{\gamma} \ll \dot{\gamma}^*$, the interaction barrier plays the dominant role, and $k_{1,1}$ increases as $U_m$ decreases. When $\dot{\gamma} \gg \dot{\gamma}^*$, on the other hand, the shear-induced aggregation takes over the dominant role, and $k_{1,1}$ increases as $\dot{\gamma}$ increases. Thus, this critical shear rate marks the transition from a slow aggregation regime with an activation delay due to non-zero potential barrier to a fast aggregation regime, with no activation barrier. In fact, if $\dot{\gamma}^*$ is also such that the pre-exponential factor is of order unity, then the resulting kinetics will be of the same order of purely Brownian diffusion-limited aggregation (DLA) in a stagnant fluid at $\dot{\gamma} \approx \dot{\gamma}^*$. Any further increase of the shear rate above $\dot{\gamma}^*$ will then produce higher coagulation rates.

The $k_{1,1}$ value given by Eq. [42] defines a characteristic aggregation time:

$$t_c \sim \frac{1}{k_{1,1}} \sim \frac{1}{\sqrt{(3\pi\alpha\mu\dot{\gamma}a^3 - U_m'')/k_B T}} e^{(U_m - 6\pi\alpha\mu\dot{\gamma}a^3)/k_B T} \tag{42}$$

An exponential dependence on $\dot{\gamma}$ for the aggregation time has been recently observed for the shear-induced aggregation of charged non-Brownian suspensions in simple shear [7]. Interestingly, the same exponential decrease with the shear rate ($\sim \exp(-\xi\dot{\gamma})$, with $\xi$ a fitting constant) of the nucleation time has been recently observed within numerical simulations of shear-induced nucleation of semi-crystalline polymers [19].

### H. Shear-driven self-accelerating aggregation kinetics

The aggregation time and the rate constant in the presence of shear display a very strong dependence upon the particle size, as is evident from Eqs. (41)-(42). For instance, in the case where the potential is fixed, the dependence on the colloid radius reads

$$t_c \sim \frac{1}{\sqrt{3\pi\alpha\mu\dot{\gamma}a^3/k_BT}} e^{-6\pi\alpha\mu\dot{\gamma}a^3/k_BT} \tag{43}$$

For example, with $\mu = 0.001$ Pa s, $\alpha = 1/3\pi$ and $\dot{\gamma} = 500\,\text{s}^{-1}$, the expression in Eq. (43) amounts to 2.26 if $a = 100$ nm and to 0.14 if $a = 200$ nm. Thus, doubling the colloid radius leads to a reduction of the characteristic time for binary aggregation by an order of magnitude.

This effect becomes of great importance when one considers the long time evolution of the coagulation process. Let us consider, e.g., a system of Brownian drops. The evolution equation for classes characterized by their size $i$ (where $i = 1, 2, 3, ..., \infty$), is given by [20]

$$\frac{dc_k}{dt} = \frac{1}{2}\sum_{i+j=k} k_{i,j}c_i c_j - c_k \sum_{j=1}^{\infty} k_{j,k} c_j \tag{44}$$

Based on the above considerations, the rate constants will be of order

$$k_{i,j} \sim \sqrt{\frac{3\pi\alpha\mu\dot{\gamma}(a_i+a_j)a_i a_j - U_m''}{k_BT}} e^{[-U_m + 6\pi\mu\dot{\gamma}(a_i+a_j)a_i a_j]/k_BT} \tag{45}$$

where the mutual diffusivity of two drops of size $i$ and $j$ respectively is defined as $D_{i,j} = k_B T(1/a_i + 1/a_j)/6\pi\mu$, leading to $Pe_{i,j} = 3\pi\mu\dot{\gamma}(a_i + a_j)a_i a_j / k_B T$. In comparison with the diffusion-limited case in stagnant fluids, where the kinetics slows down as the particles size grows by coalescence (the size dependence of the kinetics is dictated by diffusion), we can see from Eq. (45) that under shear, on the other hand, larger drops coalesce much faster, thus leading to a *self-accelerating* kinetics. In particular, at fixed $\dot{\gamma}$, an *activated* size, $a^* = (U_m / 6\pi\alpha\mu\dot{\gamma})^{1/3}$, can be defined which corresponds to a vanishing activation barrier. This may help explain the explosive rise of viscosity in non-Brownian suspensions, reported in [7], as well as in Brownian suspensions, as will be shown in the next Section, as due to self-accelerating kinetics setting in when the linear size of the growing mesoscopic structures (*clusters*, with non-coalescing colloids) reaches the activated value $a*$. The extremely rapid growth of viscosity is indeed tightly connected with the rapid growth in size which causes an increase in the volume occupied by the clusters. Associated with the increase in the effective volume fraction of clusters is the increase in (many-body) hydrodynamic interactions which are in turn responsible for the increase in the suspension viscosity.

### III. IMPLICATIONS FOR THE TIME EVOLUTION OF VISCOSITY (RHEOPEXY) IN CHARGE-STABILIZED SUSPENSIONS UNDER STEADY SHEAR

#### A. Experimental

An exponential dependence of the characteristic aggregation time upon the applied shear-rate has been observed in the shear-induced aggregation of charged suspensions [7]. Eq. (43) may provide the first theoretical justification to those evidences. However, the system in [7] was constituted of *non-Brownian* particles, hence may not be the ideal ground for a stringent test of our theory. Therefore, we carried out experiments in our lab on a system of charge-stabilized *Brownian* (polymer) particles in water, where the interplay between Brownian motion and shear is significant.

The colloidal system used to perform these experiments is a surfactant-free colloidal dispersion in water, constituted by styrene-acrylate copolymer particles supplied by BASF AG (Ludwigshafen, Germany) and produced by emulsion polymerization. The particles are nearly monodisperse and the mean radius, $a = 60 \pm 1$ nm, was characterized by both dynamic light scattering (using a Nano-ZS in Phase Analysis Light Scattering mode, Malvern, U.K.) and small-angle light scattering (using a Mastersizer 2000 instrument, Malvern, U.K). In order to avoid contamination, we performed a thorough cleaning of the suspensions by mixing with ion-exchange resins. To check that the suspensions were free of impurities after the cleaning procedure, we measured the surface tension by means of the Wilhelmy plate method with a DCAT-21 tensiometer (Dataphysics, Germany) and only suspensions exhibiting surface tension $\gtrsim 71$ mN/m have been subsequently employed in the experiments. For the shearing experiments, a small amount of electrolyte (NaCl) was added to make up the ionic background. In fact, with de-ionized suspensions, the time at which viscosity rises would be very long (~$10^1$ hours). This can seriously affect the system and the reproducibility of the experiments due to solvent evaporation. However, the final NaCl concentration in the sample (17mM) is well below the critical coagulation concentration (50mM). The stability of each suspension after adding the NaCl solution was checked by measuring the structure factor under dilute conditions, in order to make sure that the final suspension was still perfectly stable at the starting time of shearing and no aggregates were present when shearing was switched on in the rheometer. The latter is indeed a crucial point to ensure reproducibility. To induce the shear flow under shear-rate control and to *simultaneously* measure the viscosity of the flowing suspension, we used a strain-controlled ARES rheometer (Advanced Rheometric Expansion System, Rheometric Scientific). The gap between the outer cylinder and the inner one is 1 mm and the length of the latter is 34 mm. The outer cylinder is temperature controlled at $T = 298 \pm 0.1$ K and, in order to prevent evaporation, a solvent trap has been fixed on the outer rotating cylinder. For all the experiments we used deionized water (milli-Q, Millipore) and the mixing of the latex suspensions with NaCl solutions was done in such a way to avoid heterogeneities in the concentration field which would cause the aggregation kinetics to speed up in locally more concentrated zones. It is worth noting that the sampling of all the mixtures NaCl-solution/latex was done carefully with a top-cut pipette in order to avoid any local shearing during the sampling that could induce aggregation. After the shearing experiment was started, the remaining amount of (stagnant) suspension was analyzed to check the colloidal stability as a function of time under stagnant conditions. Furthermore, in order to ensure reproducibility, each time the shearing was switched on exactly 7 minutes from the time of mixing between latex and background NaCl solution.

**B. Results and discussion**

Experimental curves of viscosity as a function of time for a colloid volume fraction $\phi = 0.21$ are shown in Fig. 2 for shear rates in the Peclet range $0.49 < Pe < 0.84$. Similar curves have been obtained at volume fraction $\phi = 0.19$ and $\phi = 0.23$. For each curve in Fig. 2 a characteristic aggregation time can be extracted from the intersection of the asymptotes, as was proposed in [7]. Also in the present case, the aggregation time is an exponentially decreasing function of $\dot{\gamma}$: $t_c \sim \exp(-\xi \dot{\gamma})$, with $\xi = 0.0008$ at $\phi = 0.19$, $\xi = 0.0012$ at $\phi = 0.21$, and $\xi = 0.0018$ at $\phi = 0.23$. Using Eq. (42) with $\alpha = 1/3\pi$ we obtain $t_c \sim \exp(-0.0005\dot{\gamma})$. The more accurate result based on the numerical integration of Eq. (28), which accounts for hydrodynamic interactions in the dilute (two-body) limit, yields $t_c \sim \exp(-0.0006\dot{\gamma})$, thus not far from the approximate result from Eq. (42), where hydrodynamic interactions are neglected. A smaller $\xi$ from the theory is expected since it does not account for hydrodynamic many-body interactions which cause $\xi$ to increase with increasing colloid volume fraction [7]. A more specific study focused on the effect of many-body hydrodynamic interactions on the aggregation kinetics in shear will appear elsewhere.

## IV. CONCLUSION

Starting from the two-body Smoluchowski equation for interacting Brownian particles under shear, we derived an approximate theory for the irreversible aggregation kinetics of colloids in linear flows. The predictions, with no fitting parameter, are in good agreement with numerical data, up to Peclet $\sim 10^2$. With interaction barrier (as for e.g. DLVO-interacting systems), a rate-equation for the kinetic constant of aggregation has been derived which appears to be in Arrhenius form and consists of a frequency factor ($\propto \dot{\gamma}^{-1/2}$) multiplying an exponential, $e^{-(U_m - 6\pi\alpha\mu\dot{\gamma}a^3)/k_B T}$, giving the probability (efficiency) of sticking upon collision. The shear rate $\dot{\gamma}$ is effective in diminishing the activation barrier $U_m$, and the aggregation kinetics rises by orders of magnitude just above an activated value of $\dot{\gamma}$ which causes the barrier to vanish. Our theory, extending activated rate-theory of Brownian particles [21] to activated barrier-crossing driven by shear, explains the induction period followed by self-accelerating aggregation kinetics and explosive rise of suspension viscosity observed in many sheared complex and biological fluids [2,7], including protein solutions [4,5]. This paves the way for a microscopic, possibly quantitative, understanding of *shear-induced* structure-formation processes [3] and phase transitions (e.g. gelation) in soft matter systems.

**ACKNOWLEDGEMENTS**

Financial support from the Swiss National Science Foundation (grant. No. 200020-126487/1) is gratefully acknowledged. Discussions with Dr. M. Lattuada, Dr. E. Del Gado, and Dr. V. Guida are gratefully acknowledged.

REFERENCES

[1] P. Coussot, *Rheometry of Pastes, Suspensions, and Granular materials: Applications in Industry and Environment* (Wiley, New York, 2005).

[2] P. Coussot, Q. D. Nguyen, H. T. Huynh, and D. Bonn , J. Rheol. **46**, 573 (2002).

[3] J. Vermant and M. J. Solomon J. Phys: Cond. Matter **17**, R187-R216 (2005).

[4] K. M. N. Oates *et al.*, Journal of the Royal Society-Interface **3**, 167-174 (2006).

[5] S. Rammensee, U. Slotta, T. Scheibel, and A. R. Bausch, Proc. Natl. Acad. Sci. U.S.A. **105**, 18 (2008).

[6] H. J. Jin and D. L. Kaplan, Nature **424**, 1057-1061 (2003).

[7] J. Guery, E. Bertrand, C. Rouzeau, P. Levitz, D. A. Weitz, and J. Bibette, Phys. Rev. Lett. **96**, 198301 (2006).

[8] V.G. Levich, *Physicochemical Hydrodynamics* (Prentice-Hall, Englewood Cliffs, NJ, 1962).

[9] E. J. W. Verwey and J. Th. G. Overbeek, *Theory of the Stability of Lyophobic Colloids* (Elsevier, New York, 1948).

[10] J. K. G. Dhont, J. Fluid Mech. **204**, 421 (1989).

[11] B. J. Ackerson and N. A. Clark, Physica A **118**, 221-249 (1983).

[12] J. Bergenholtz, J. F. Brady and M. Vicic, J. Fluid Mech. **456**, 239 (2002).

[13] S. Melis, M. Verduyn, G. Storti, M. Morbidelli, and J. Bałdyga, AIChE J. **45**, 7 (1999).

[14] R. A. Lionberger, J. Rheol. **42**, 843 (1998).

[15] G. K. Batchelor and J. T. Green, J. Fluid Mech. **56**, 401 (1972).

[16] G. I. Barenblatt, *Scaling, Self-similarity, and Intermediate Asymptotics*, pp. 39-43 (Cambridge University Press, Cambridge, 1996).

[17] H. Masliyah and S. Bhattacharjee, *Electrokinetic and Transport Phenomena*, pp. 446-447 (Wiley, Hoboken NJ, 2006).

[18] H. Risken, *The Fokker-Planck Equation* (Springer, Berlin, 1996).

[19] R. S. Graham and P. D. Olmsted, preprint arXiv:0909.0213 (2009).

[20] S. Chandrasekhar, Rev. Mod. Phys. **15**, 1 (1943).

[21] H. A. Kramers, Physica **7**, 284-304 (1940).

FIGURE 1

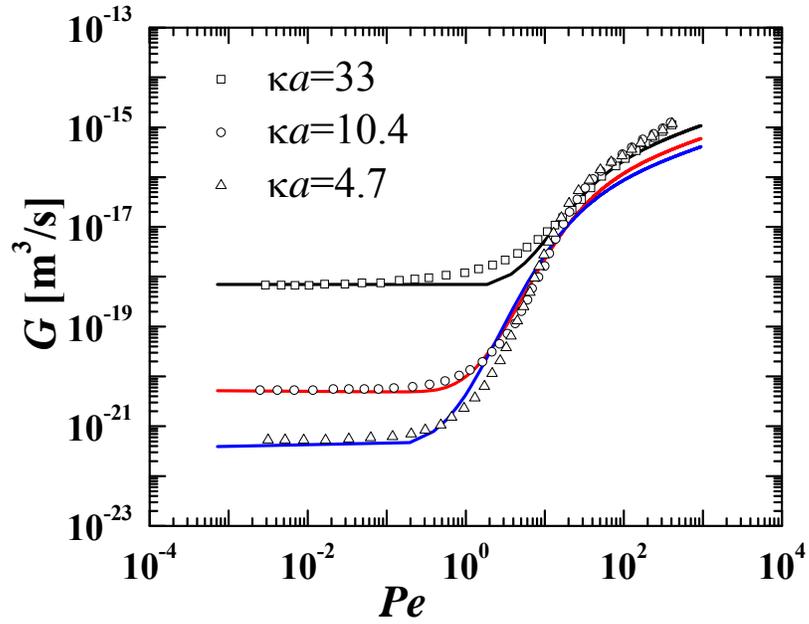

FIG. 1. (Color online) Comparison between calculations based on the proposed theory (Eq. (23)) and the numerical simulations of the full convective diffusion equation (Eq.(3)) reported in [13] for three different ionic strength conditions. The colloid surface potential is fixed and equals -14.7 mV.

FIGURE 2

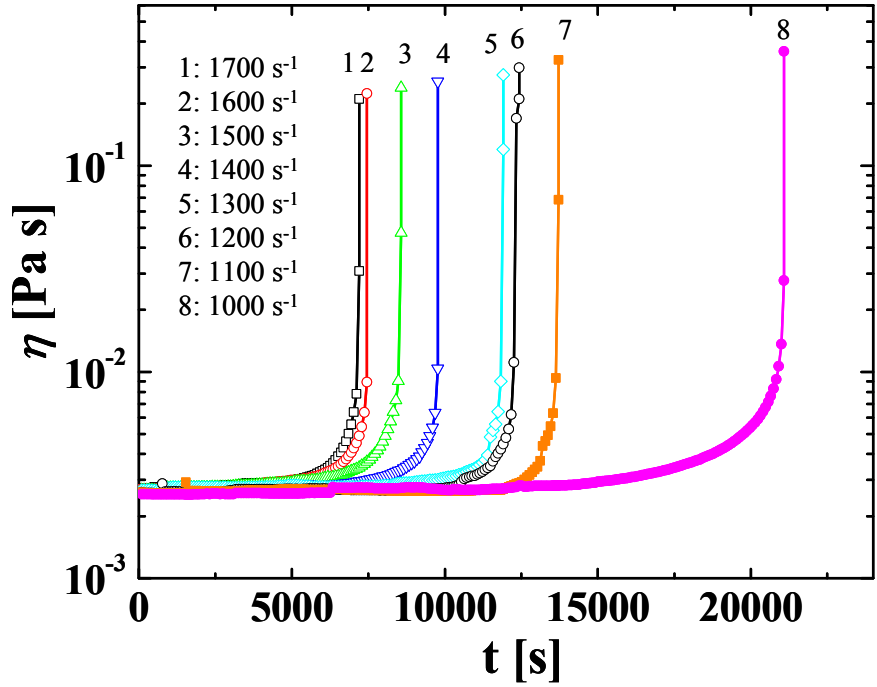

FIG. 2. (Color online) Viscosity as a function of time under steady shear for charge-stabilized polystyrene colloids ($\phi = 0.21$) for different shear rate values, $\dot{\gamma}$ (see legend). The NaCl concentration is 17mM for all cases. The characteristic time of aggregation can be estimated as shown in [7].